\documentclass[sigconf]{acmart}

\usepackage{todonotes}
\usepackage{hyperref}
\usepackage{booktabs}
\usepackage{diagbox}
\usepackage[frozencache=true,cachedir=minted-cache]{minted} 
\usepackage[T1]{fontenc} 
\usepackage{sourcecodepro} 

\definecolor{myred}{rgb}{1, 0.941, 0.941}
\definecolor{mygreen}{rgb}{0.941, 0.9725, 0.949}
\definecolor{myyellow}{rgb}{1, 0.988, 0.941}

\setminted{fontsize=\small}
\setminted{autogobble=true}

\def\codeinline#1{\mintinline{text}{#1}}

\def\mllint{\codeinline{mllint}\,}

\AtBeginDocument{%
  \providecommand\BibTeX{{%
    \normalfont B\kern-0.5em{\scshape i\kern-0.25em b}\kern-0.8em\TeX}}}

\setcopyright{acmcopyright}
\copyrightyear{2021}
\acmYear{2021}

\acmConference[ICSE 2022 -- SEIP]{ICSE 2022 - Software Engineering in Practice}{May 21--29, 2022}{Pittsburgh, PA, USA}
\acmPrice{0.00}

\begin{document}

\settopmatter{authorsperrow=4}

\title{``\textit{Project smells}'' --- Experiences in Analysing the Software Quality of ML Projects with \texttt{mllint}}

\author{Bart van Oort}
\affiliation{%
  \institution{TU Delft}
  \institution{AI for Fintech Research, ING}
  \city{Delft}
  \country{Netherlands}
}
\email{bart.van.oort@ing.com}

\author{Lu\'is Cruz}
\affiliation{%
  \institution{TU Delft}
  \city{Delft}
  \country{Netherlands}
}
\email{l.cruz@tudelft.nl}

\author{Babak Loni}
\affiliation{%
  \institution{ML Engineering Chapter, ING}
  \city{Amsterdam}
  \country{Netherlands}
}
\email{babak.loni@ing.com}

\author{Arie van Deursen}
\affiliation{%
  \institution{TU Delft}
  \city{Delft}
  \country{Netherlands}
}
\email{arie.vandeursen@tudelft.nl}


\begin{abstract}
Machine Learning (ML) projects incur novel challenges in their development and productionisation over traditional software applications, though established principles and best practices in ensuring the project's software quality still apply. While using static analysis to catch code smells has been shown to improve software quality attributes, it is only a small piece of the software quality puzzle, especially in the case of ML projects given their additional challenges and lower degree of Software Engineering (SE) experience in the data scientists that develop them. We introduce the novel concept of \textit{project smells} which consider deficits in project management as a more holistic perspective on software quality in ML projects. An open-source static analysis tool \mllint was also implemented to help detect and mitigate these. Our research evaluates this novel concept of project smells in the industrial context of ING, a global bank and large software- and data-intensive organisation. We also investigate the perceived importance of these project smells for proof-of-concept versus production-ready ML projects, as well as the perceived obstructions and benefits to using static analysis tools such as \mllint. Our findings indicate a need for \textit{context-aware} static analysis tools, that fit the needs of the project at its current stage of development, while requiring minimal configuration effort from the user.
\end{abstract}

\keywords{project smells, software quality, machine learning, mllint, code smells, context-aware, static analysis, dependency management, Python}

\renewcommand\copyright{{\textcopyright}}

\maketitle


\section{Introduction} \label{sec:introduction}

The ubiquity of Machine Learning (ML) and Artificial Intelligence (AI) solutions to complex computing problems demands development processes to help transform a proof-of-concept ML experiment into a well-engineered ML application, running continuously in a production environment \cite{amershi2019-se4ml-case,menzies2020lawsofSEforAI,serban2021ml-architecture,haakman2020ai}. These development processes on the one hand incorporate novel ideas to deal with the challenges that developing ML applications poses over developing traditional software applications, but on the other hand also include established Software Engineering (SE) best practices. After all, quoting \citet{menzies2020ai-effect}, ``An AI system is a software-intensive system, and the established principles of designing and deploying quality software systems that meet their mission goals on time still apply''.

However, productionising is difficult, especially in the case of ML systems given their additional challenges, such as data management, testing and reproducibility \cite{kriens2019software,nascimento2020se4ml-slr}. Traditional software engineering historically struggled with this too, but has seen the implementation of a host of tools to help with productionisation in various stages of the software development lifecycle.
For example, using static analysis to enforce best practices and catch code smells, helps catch bugs earlier and improve software quality attributes, such as reliability, maintainability and reproducibility \cite{lacerda2020smellsSLR}.

Especially in ML projects, code smells are only a small piece in the software quality puzzle. We noticed this first-hand in our previous research on the prevalence in code smells in open-source ML projects: nearly half of the analysed projects struggled with managing their code dependencies \cite{bvo-wain2021}. We realised that a more holistic approach to code smells, `\textit{project smells}', would be required.

To the end of automatically detecting such project smells and giving practical advice on how to fix those, we implemented \mllint. \mllint\footnote{\url{https://github.com/bvobart/mllint}} is an open-source command-line utility to evaluate the software quality of Python ML projects by performing static analysis on the project's source code, data and configuration of supporting tools. \mllint aims to help ML practitioners in developing and maintaining production-grade ML and AI projects.

We argue that many data-driven companies may benefit from an SE for ML tool such as \mllint. This research gauges how well these project smells as detected by \mllint fit the context of ML development at our industrial partner ING. ING is a global bank and large software- and data-intensive organisation with a strong European base that offers retail and wholesale banking services to 38.5 million customers in over 40 countries \cite{ing-at-a-glance} and has 15.000 employees in IT, software and data technology \cite{ing-afr-lab}. ING has extensive use-cases for increasing its business value with AI and ML, such as assessing credit risk, fighting economic crime by monitoring transactions and improving customer service. As part of a major shift in the organisation to adopt AI and ML and become data-driven, ING is defining standards for the different processes around the lifecycle of ML applications \cite{haakman2020ai}.

To measure the fit of project smells in this context, we qualitatively analyse the reports generated by \mllint on ING projects and combine them with feedback from ML practitioners. Additionally, we ask practitioners to run \mllint on their projects and provide us feedback on their experiences with \mllint and its concepts. By doing so, we aim to uncover the obstacles of implementing specific best practices, as well as the perceived benefits and drawbacks of using static analysis tools such as \mllint to verify SE practices in ML projects. Additionally, we investigate how ML practitioners perceive the importance of \mllint's linting rules on proof-of-concept versus production-ready projects, as the former may not require as rigorous software quality checks as the latter do.

More formally, our research questions are as follows:


\begin{description}
    \item[RQ1] How do the project smells as detected by \mllint fit the industrial context of a large software- and data-intensive organisation like ING?
    %

    \item[RQ2] What differences do ML practitioners perceive in the importance of \mllint's linting rules between proof-of-concept and production-ready projects?

    \item[RQ3] What are the main obstacles for ML practitioners towards implementing specific best practices?

    \item[RQ4] What are the perceived benefits of using static analysis tools such as \mllint to verify SE practices in ML projects?
\end{description}

The rest of this paper is structured as follows. Section~\ref{sec:background} describes influential research in the field of SE for ML that supports this research. Section~\ref{sec:mllint} elucidates the concept of \textit{project smells} and \mllint, detailing two major challenges in its development. Section~\ref{sec:methodology} explains the methodology used to answer our research questions. In Section~\ref{sec:results}, we present the findings from applying our methodology and answer our research questions. Section \ref{sec:discussion} then combines and discusses these findings along the themes of version controlling data, dependency management and static analysis tool adoption. We then discuss the threats to the validity of our research in Section~\ref{sec:threats} and conclude with future work in Section~\ref{sec:conclusion}.

The contributions of this research are as follows:
\begin{itemize}
    \item The novel concept of \textit{project smells} as a holistic perspective on software quality in ML projects.
    \item An open-source static analysis tool \mllint\footnote{\url{https://github.com/bvobart/mllint}} to help with detecting and mitigating these project smells.
    \item Experiences, insights and perceptions on project smells in an industrial context.
\end{itemize}






\section{Background} \label{sec:background}

Both SE and ML are well studied in literature, though their intersection is still an emerging field of research \cite{amershi2019-se4ml-case,menzies2020lawsofSEforAI,nascimento2020se4ml-slr}.

\citet{sculley2015hidden} were among the first to investigate risk factors in the design of real-world ML systems at Google through the lens of technical debt. In doing so, they unearthed several anti-patterns in ML system design, including glue code---the tendency for ML applications to consist of code that glues together functionalities from various general-purpose libraries---and configuration debt---the tendency for both researchers and engineers to see configuration and configurability of the ML application as an afterthought \cite{sculley2015hidden}. Continued research at Google investigating production-readiness and the reduction of technical debt in ML systems, resulted in ``\textit{The ML Test Score}'' \cite{google2017production-rubric}: a rubric with 28 specific tests and monitoring needs, along with a scoring system to determine the production-readiness of ML systems. There are four categories, each with seven tests: \textit{Data}, \textit{Model}, \textit{Infrastructure} and \textit{Monitoring}. Executing a test manually, documenting and distributing the results, earns the project half a point. A full point is awarded if that test is automated and runs regularly. The awarded points are then summed up within each category and the lowest of these sums is the final production-readiness score. A score between 3 and 5 is interpreted as ``\textit{Strong levels of automated testing and monitoring, appropriate for mission-critical systems.}'' \cite{google2017production-rubric} 

\citet{amershi2019-se4ml-case} at Microsoft also used experiences from engineering ML applications in their case study. Their study resulted in several best practices and three aspects of engineering ML / AI applications that make them fundamentally different from traditional software applications. One aspect is the discovery and management of data: ML applications also need to deal with finding, collecting, cleaning, curating and processing their input data. This data also needs to be stored and versioned, for which in contrast to code there were no well-designed technologies to do so \cite{amershi2019-se4ml-case}. Another challenge is the customisation and reuse of ML models on problems in different domains or with slightly different input formats, as this may require retraining or even replacing the model with new or additional training data. Finally, strict modularity between ML models is difficult to achieve, as models are not easily extensible and multiple models may interact with each other in unexpected ways \cite{amershi2019-se4ml-case}. \citet{kriens2019software} recognise this and propose a partial solution in the form of OSGi-like metadata for ML models.

The aforementioned challenges are reflected in systematic literature reviews (SLRs) such as \cite{nascimento2020se4ml-slr}, \cite{washizaki2019se-patterns} and \cite{alamin2021mlsa-quality-assurance}. \citet{nascimento2020se4ml-slr} analysed the limitations and open challenges found in the SE for ML field of research, noting that testing, AI software quality and data management are three of the main challenges faced by professionals in the field. They also report on several SE practices, approaches and tools for dealing with these challenges. On the topic of testing ML systems, \citet{zhang2020ml-testing} performed an extensive SLR of various techniques to do this.
\citet{washizaki2019se-patterns} similarly performed an SLR on SE design patterns for ML techniques, identifying several good and bad patterns for engineering ML software. \citet{muralidhar2021mlops-antipatterns} also identify MLOps anti-patterns.
More recently, \citet{alamin2021mlsa-quality-assurance} conducted an in-depth literature review, resulting in a taxonomy of different quality assurance challenges for ML software applications, which includes dealing with data dependencies and ML-specific technical debt. \citet{bogner2021-slr-ai-tech-debt} further investigates technical debt in ML systems, identifying new forms of such debt, 72 anti-patterns (most of them relating to models and data) and 46 potential solutions to them.


Our research builds on the work of SE4ML \cite{se4ml-website}, who have identified 45 best practices for engineering trustworthy ML applications \cite{serban2020se4ml-adoption,serban2021se4ml-practices,blom2021automl}. They also measured the adoption of these best practices, both in academic and industry use \cite{serban2020se4ml-adoption}. Their findings indicate that larger teams tend to adopt more best practices and that traditional software engineering practices tend to have a lower adoption than ML-specific best practices. More recently, they also studied challenges and solutions in an SLR about software architecture for systems with ML components \cite{serban2021ml-architecture}. Along with new ML-specific challenges, they also found that traditional software architecture challenges also play an important role in architecting ML systems.

Lastly, in our previous work, we analysed the prevalence of code smells in ML projects \cite{bvo-wain2021}. Aside from widespread code duplication in ML projects and several false positives in Pylint, we coincidentally found that nearly half of the projects we analysed struggled with dependency management, so much so that manual adjustments were needed to allow error-free installation of the Python libraries that they used. This severely hurts the maintainability and reproducibility of these projects.

\section{mllint} \label{sec:mllint}

Following from the related work, a pattern emerges suggesting that code smells are only a small factor in the software quality of ML applications. Code smells do ``\textit{have a strong relationship with quality attributes, i.e., with understandability, maintainability, testability, complexity, functionality, and reusability}'' \cite{lacerda2020smellsSLR}, but they do not paint the complete picture, especially in ML given the extra challenges in their development over traditional software applications.


Thus, to more accurately assess the software quality of ML applications, a more holistic approach would be required, where instead of \textit{code} smells, we analyse \textit{project} smells. Such project smells are concerned with deficits in how an ML project is managed, including poor dependency management (as outlined in \cite{bvo-wain2021}), lack of version control for code or data, unit testing, proper Continuous Integration (CI) configurations, or effective static analysis tooling. Code smells are also a subcategory of project smells.

To the end of automatically detecting such project smells and giving practical advice on how to fix those, we implemented \mllint. \mllint is a command-line utility to evaluate the software quality of ML projects written in Python by performing static analysis on the project's source code, data and configuration of supporting tools. The aim of \mllint is threefold. First, \mllint aims to help data scientists and ML engineers in creating and maintaining production-grade ML and AI projects, both on their own personal computers as well as on CI. Secondly, it aims to help ML practitioners inexperienced with SE techniques explore and make effective use of battle-hardened SE for ML tools in the Python ecosystem, Finally, \mllint aims to help ML project managers assess the quality of their ML projects and receive recommendations on what aspects of their projects they should focus on improving.

\subsection{Implementation} \label{sec:implementation}

\mllint analyses a project with linting rules in five categories, which roughly correspond to the project smells as previously outlined. These categories are based on well-known SE practices in traditional software application development, as well as SE for ML best practices from (grey) literature, such as SE4ML's collection of best practices \cite{se4ml-website,serban2021se4ml-practices} and Google's Rules for ML \cite{google-rules-of-ml}. Each category is described as follows. 

\begin{description}
    \item[Version Control] This category comprises both version controlling source code (with Git), as well as version controlling data. The latter is particularly relevant to ML applications.
    \item[Dependency Management] This category entails checking whether the project manages its code dependencies (e.g. used libraries) in a reproducible and maintainable manner, to mitigate the dependency management issues found in \cite{bvo-wain2021}.
    \item[Continuous Integration] The rule in this category checks whether the project has a CI configuration file. 
    \item[Code Quality] This category is concerned with code smells and runs a set of linters (Pylint, Mypy, Black, \codeinline{isort} and Bandit) to detect and help mitigate them.\footnote{This category could be extended with tools for detecting ML-specific code smells, such as \texttt{dslinter} \cite{haakman2020thesis}.}
    \item[Testing] This category analyses testing practices in the project by counting the number of test files, the number of tests passed and the test coverage. Since \mllint performs \textit{static} analysis, it will not run the tests, but instead expects a test- and coverage report from a prior test run.
\end{description}

Each category contains linting rules that analyse and score how the best practice referred to by the category is implemented in the project. For example, the \textit{Version Control} category contains rules such as ``\textit{Project uses Git}'', ``\textit{Project should not have any large files in its Git history}'' and ``\textit{Project uses Data Version Control}''. The checks imposed by these linting rules are based upon prevalent tooling and usage techniques found in the industry.

\begin{sloppypar}
Additionally, users can define custom rules in their \mllint configuration by referring to a custom program that performs this custom check and returns the resulting score and corresponding details. This allows users to implement their own linting rules for verifying internal team or company practices. Additionally, it can help to prototype new rules for \mllint before they are included in \mllint's core set of linting rules.
\end{sloppypar}

\begin{figure}[htbp]
    \centering
    \includegraphics[width=0.9\linewidth]{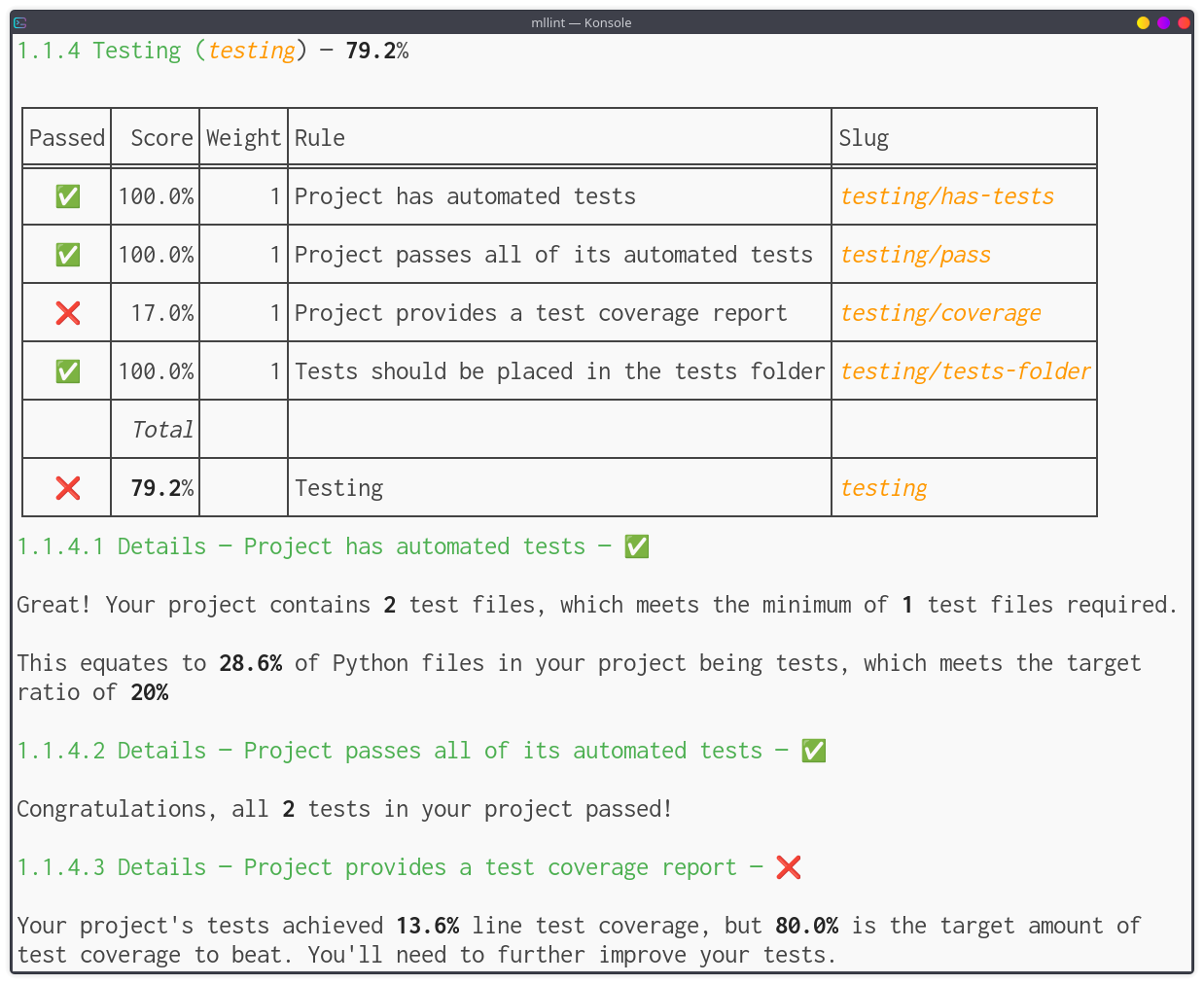}
    \vspace{-1ex}
    \caption{Example snippet from an \mllint report rendered to the terminal. The full report can be found on GitHub.\protect\footnotemark}
    \label{fig:mllint-report}
\end{figure}
\footnotetext{\url{https://github.com/bvobart/mllint/blob/main/docs/example-report.md}}

After its analysis, \mllint outputs a Markdown-formatted report that is by default pretty-printed to the terminal. This report contains a score for each rule (between 0 and 100\%), often along with details that explain the score, provide extra information derived from the analysis and / or provide recommendations on how to make the rule pass. For an example of such a terminal-rendered \mllint report, see \autoref{fig:mllint-report}.

The experienced practitioner might note that \mllint in its current state primarily focuses on project smells that are also applicable to non-ML projects, but has few rules that specifically apply only to ML projects. There are two reasons for this: first, such more general SE tools and techniques (e.g. dependency management and linting for code quality) are more alien to data scientists than ML-specific tools, given the low degree of SE experience in data scientists. Secondly and more practically, given our limited amount of development resources for \mllint, we have yet to implement more ML-specific linting rules.

In the following two sections, we explain two of the challenges faced in architecting \mllint and its rules, along with how we approached them. Note that \mllint is currently a research prototype. There are still many linting rules to implement and a host of ML project tools and architectures to support. As such, \mllint is yet to reach its full potential, though it does pave the way towards using static analysis techniques to improve the quality of ML projects. \mllint is built with an extensible architecture so that the list of supported practices can continuously be extended.

\subsection{Challenge 1: Mapping high-level best practices to practical guidelines}
The SE for ML best practices found in academic sources such as \cite{se4ml-website,serban2021se4ml-practices} tend to be quite high-level: they explain a concept or technique for a project to adhere to, but often provide little direct, practical recommendations on how to implement it correctly. An example of this is the best practice to use static analysis tools for checking code quality\footnote{\url{https://se-ml.github.io/best_practices/03-use_static_analysis/}}, which does not recommend any specific linting tools to employ or what kinds of linting rules to enable---in part, to remain timeless and general. But especially within the plethora of language-supporting, supplementary tools and libraries that exists within the Python ecosystem, it can be very difficult and time-consuming to find the right tools or configuration.

The aim of \mllint is therefore to give practical advice to its users; concrete tools, techniques and guidelines that the user can implement such that the SE for ML best practices are fulfilled, along with automated checks to detect the degree of adoption. However, figuring out which exact tools to advocate for implementing which best practice is non-trivial, especially as programming ecosystems change. This is best done by looking at what tools are prevalent and popular in the industry.

Additionally, \mllint is an automated, command-line tool, so based only on the project's source code (e.g. the contents of its Git repository), it has to be able to reliably computationally check whether the project adheres to each practice. This makes it difficult, in some cases impossible, to verify whether certain team- or governance-related best practices are being upheld. Two examples of this are ``\textit{Establish Responsible AI Values}'' and ``\textit{Perform Risk Assessments}'' \cite{se4ml-website}. These are team or company processes that a source code analysis tool such as \mllint will not be able to enforce.

To find appropriate linting rules and practical advice, we started by exploring the practical side of the SE for ML landscape and how their best practices relate to practical implementations. This was done by estimating the measurability of the best practices from SE4ML \cite{se4ml-website,serban2021se4ml-practices} and Google's Rules for ML \cite{google-rules-of-ml}. For each practice, we explored ways to detect adherence to it in the source code of an ML project, how reliable such an approach would be and, by extension, how feasible it would be to reliably and accurately measure adherence to this practice. Our indication of measurability was given as one of five colours between red (not measurable), yellow (technically measurable, but likely to be unreliable or inaccurate) and green (measurable in a reliable and accurate way). 


As an example, consider the best practice to use Continuous Integration \cite{se4ml-website}. This was marked yellow, as it is easily possible to detect whether a project has a CI configuration in its repository --and CI configurations are also machine-readable-- but it is difficult to determine whether this configuration contains an appropriate set of CI jobs for the project. By contrast, the best practice ``\textit{Check that Input Data is Complete, Balanced and Well Distributed}'' \cite{se4ml-website} was marked green, since this data should be available through the software repository and only requires a few statistical checks on the data, possibly through tools like GreatExpectations\footnote{\url{https://greatexpectations.io/}} or TensorFlow Data Validation\footnote{\url{https://github.com/tensorflow/data-validation}}. Finally, the aforementioned best practice to ``\textit{Establish Responsible AI Values}'' was marked red, as this is a team / organisational value that cannot be deduced from the project's software repository.

After this measurability analysis, we simply picked the low-hanging fruits, i.e., the most measurable, yet also easy to implement best practices to become our first best practices. An iterative approach was then taken in constant collaboration with experienced ML engineers from ING to determine which best practices were most useful next.


\subsection{Challenge 2: Heterogeneity of ML projects}
Another big challenge for \mllint comes from the many different kinds of ML projects. An ML project could be plain-old Python using basic ML libraries, but could also be based on a framework like TensorFlow or PyTorch, for each of which a different project architecture and tooling might be preferable. If a company has their own ML infrastructure or platform, then this could also impose different requirements to the project's layout and tooling. Furthermore, if the company uses proprietary tools for fulfilling certain practices, \mllint may not recognise them and will not be able to assess whether the best practice is followed correctly, resulting in incorrect recommendations. All in all, some linting rules may not make sense on certain projects, or need to adapt what they recommend for different kinds of projects.

Aside from the technical differences, ML projects may also differ in maturity. An ML project that is only a proof of concept does not need to be as highly engineered, reproducible and maintainable as a production-ready or fully productionised project. Surely, it should get its basics right, as the best practice ``\textit{Keep the first model simple and get the infrastructure right}'' (rule \#4 of Google's Rules of ML \cite{google-rules-of-ml}) also endorses, but as an example, it may not be worthwhile fixing all linter warnings or achieving full test coverage. The more mature the project, the more important these engineering principles become though. Since \mllint's recommendations may steer the engineering process, \mllint should account for differences in maturity, by adjusting the weights of its rules to match what is important to the project at the current stage of development. This paper therefore also investigates the perceived differences in the prioritisation of each of \mllint's rules.

Finally, there will always be tools, techniques and practices that \mllint will not recognise or have linting rules for, such as proprietary tools and internal company / team practices. To provide some degree of support for such cases, \mllint allows users to define custom rules in its configuration that run some arbitrary script or program to score and provide recommendations on a custom practice. Such custom rules also provide a testing ground for new linting rules that may later be published as a plugin to \mllint, or even be built into \mllint.

Summarising, the challenge of heterogeneity of ML projects to tools like \mllint is still an open challenge. However, our proposals to solving it may at least limit its impact. These include configurability of enabled rules (with sensible defaults), automatic adaptation of linting rules to different kinds of technology stacks present in projects, and custom linting rules.


\begin{figure*}[hbtp]
    \centering
    \includegraphics[width=0.8\textwidth]{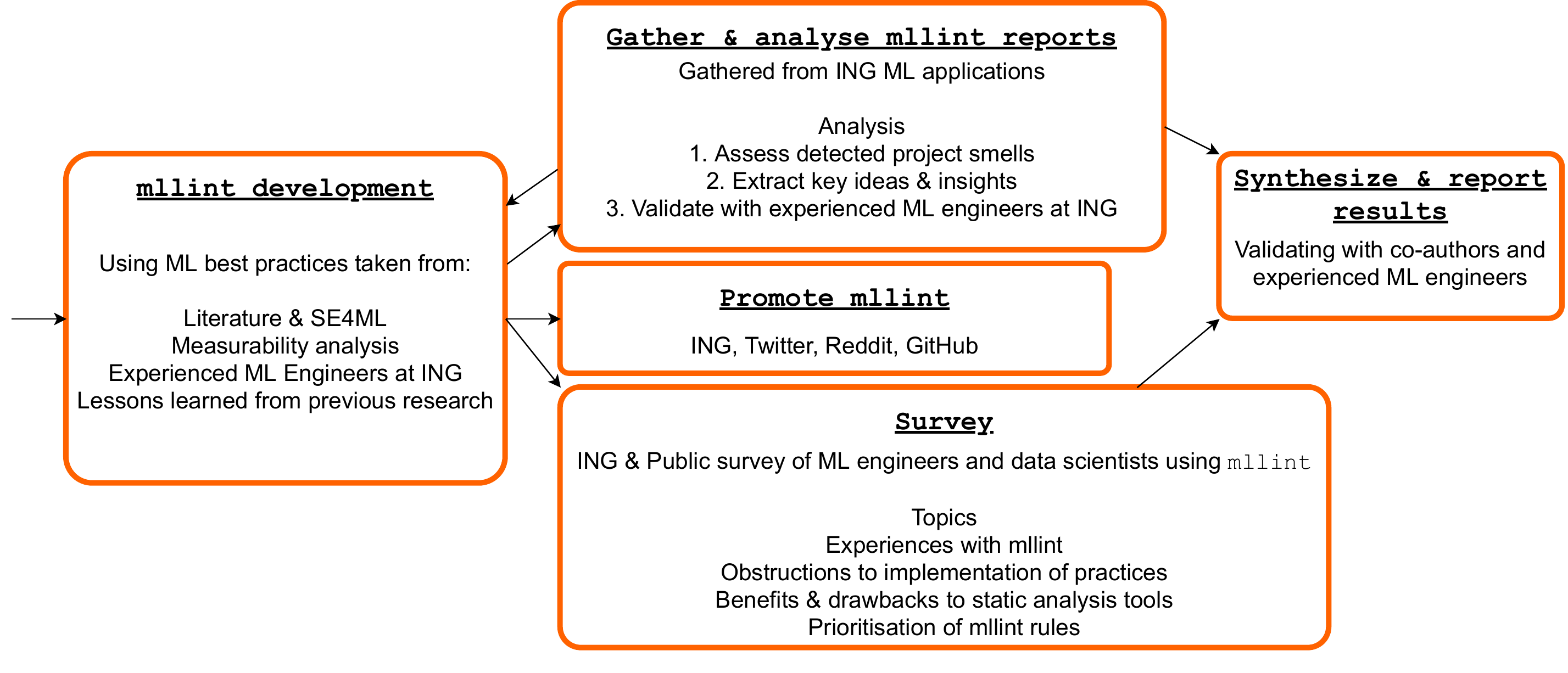}
    \vspace{-1.5em}
    \caption{Overview diagram of the methodology used in this paper.}
    \vspace{-1em}
    \label{fig:methodology}
\end{figure*}

\section{Methodology} \label{sec:methodology}

To answer the research questions posed in the introduction, we employed a mixed-methods approach. An overview of this is displayed in \autoref{fig:methodology}. First, we gathered and qualitatively analysed the \mllint reports of eight ML projects at ING. Secondly, we asked and encouraged ML practitioners from ING and open-source communities alike, to try \mllint on one or more of their projects and evaluate the reports that it produced. Subsequently, we ran a survey with 22 users of \mllint to evaluate the efficacy of the tool, as well as gather insights on how users prioritise each of the implemented rules. We also used insights from informal, open-ended interviews performed with ML practitioners within ING, partially instigated by a desire for deeper elaboration on some of the survey's answers.\footnote{While the survey was anonymous, the participant could fill in an email address for us to contact if answers were found to be either unclear or particularly interesting.}

\subsection{Qualitative analysis of \mllint reports}

To help answer RQ1, we gathered and qualitatively analysed the reports that \mllint generated for eight ML applications within ING. For seven of these projects we got access to the source code and ran \mllint on it ourselves. For the other, we asked one of its developers to run \mllint on their project themselves and forward us the report, which we then discussed with the developer.

The analysis of the reports then consists of three stages. First, we manually browse through each report to inspect the scores and details for each linting rule, assessing what project smells were detected and how deeply ingrained in the project they are. Are they simple oversights during development? Are they \mllint false positives? Were the developers unaware of the advocated best practice so far? Does the project show evidence of them applying the best practice, but in way different to what \mllint recognises, or did the developers actively choose not to implement the best practice being advocated by the rule? Whenever uncertain, we ask the project's developers for further clarification and verification of these ideas.

Secondly, combining the insights taken from multiple reports, we deduce patterns in the prevalence of the detected project smells. Which project smells are most and least often detected? Are there any project smells that are systemically ignored, or do the developers have suitable alternatives for these practices? This results in a list of key ideas and insights about \mllint's project smells.

Finally, to validate the findings and gauge the significance of the key ideas taken from stage 2 within the context of ING, we discuss the findings and key ideas with experienced ML engineers at ING.



\subsection{Survey}

Based on our research questions, we designed a survey\footnote{Survey available online: \url{https://doi.org/10.6084/m9.figshare.18777821.v1}.} to evaluate the efficacy of \mllint as a tool and gather insights on the prioritisation of each rules. The survey starts by gathering demographics such as the participant's profession, the type of organisation they work in, their team size \& composition and their experience in the fields of SE and ML. We then asked participants about their experiences with \mllint by asking about their first impressions, the amount of \mllint recommendations they have or would apply to their ML projects, and how much they agree or disagree with a few statements on how helpful the descriptions of \mllint's rules are and whether they would consider employing \mllint in their ML project development and code review process.

Next, in separate questions, we ask what benefits and drawbacks static analysis tools such as \mllint have for the participant in validating SE practices for ML projects. We use this to answer RQ4 and RQ3 respectively. We also ask questions about rules that the user disabled or was not able to implement and what was obstructing them in doing so, which we also use for answering RQ3. Furthermore, we ask participants what features or rules they think \mllint is still missing and what features they think could be improved.

In the final part of the survey, to answer RQ2, we ask participants to rate the importance of each of the linting rules currently implemented in \mllint. Possible answers are on a Likert-scale, ranging from `\textit{Not important}' to `\textit{Absolutely Essential}', with the addition of an `\textit{I don't know}' option. Since the priority of each rule may be different in different stages of the lifecycle of an ML project, we ask the participant to do this for both a \textit{proof-of-concept} and a \textit{production-ready} ML project. Since it may not be entirely clear what these terms entail, we provide the user with the following definitions:

\vspace{-1ex}
\begin{description}
    \item[Proof-of-concept project] A project that primarily serves as an example to show that the concept of the project works and will scale. Imagine that this is to show supervisor that it is worthwhile to further develop this project into one that can eventually be deployed to the production environment.
    \item[Production-ready project] A project that is mature enough to be deployed to the production environment (or already is). This requires rigorous project quality standards, such that the application is stable and will behave as expected.
\end{description}

For open-ended questions, we codified each of the answers by analysing each answer, sentence by sentence, marking the topics that they discuss and denoting their sentiment towards it; are they being positive or negative, or listing advantages or caveats to take into account? We also took note of any specific, insightful remarks from the answers. As an example, the phrase ``\textit{\mllint provides a good checklist of things to do to improve ML project quality}'' would be marked as having a positive sentiment and tagged with `\textit{quality checklist}', `\textit{project quality}' and `\textit{guides planning}'.

The survey was spread among ML practitioners at ING through their data science and data engineering mailing lists and Slack channels and the AI for FinTech Research Lab. It was also presented at ING's ML engineering chapters and two workshop sessions were held at ING Analytics, one live and one pre-recorded. Furthermore, we publicised \mllint and a public copy of the survey through our academic network, a tweet and a Reddit post. The discussion that the Reddit post triggered, as well as notes from open-ended discussions conducted with experienced ML engineers from ING after the survey, were codified similarly to the survey answers and used to answer RQ3 and RQ4.

In total, 22 people filled in our survey, most of them ML engineers, of which one chapter lead ML Engineering, two chapter leads Data Science, two ML researchers and two PhD students. 14 participants work at ING, 4 at some other non-tech company and 4 at a university or non-commercial research lab. On average, participants had between 2 to 6 years of experience at developing ML applications and between 4 to 10 years of experience in Software Engineering (defined in the survey as ``\textit{designing, implementing, testing and maintaining complex software applications}''). They tend to work in teams of 6 to 9 members, of which on average 4 have a strong background in SE.

Most participants noted they had used \mllint on one project at the time of filling in the survey, only three participants had run \mllint on two to five projects. Impressions are overall positive, with participants noting that the terminal interface is pretty and that the reports are well-organised, as well as ``\textit{informative to people unfamiliar with ML tooling and/or Python workflows}''. One participant noted the tool ``\textit{should be a standard on ML projects}''. However, participants also noted that \mllint is still early in development and that some were overwhelmed with the amount of terminal output, especially in the presence of many code quality linter messages.

\begin{figure}[h]
    \centering
    \vspace{-0.75em}
    
    \includegraphics[width=0.8\linewidth]{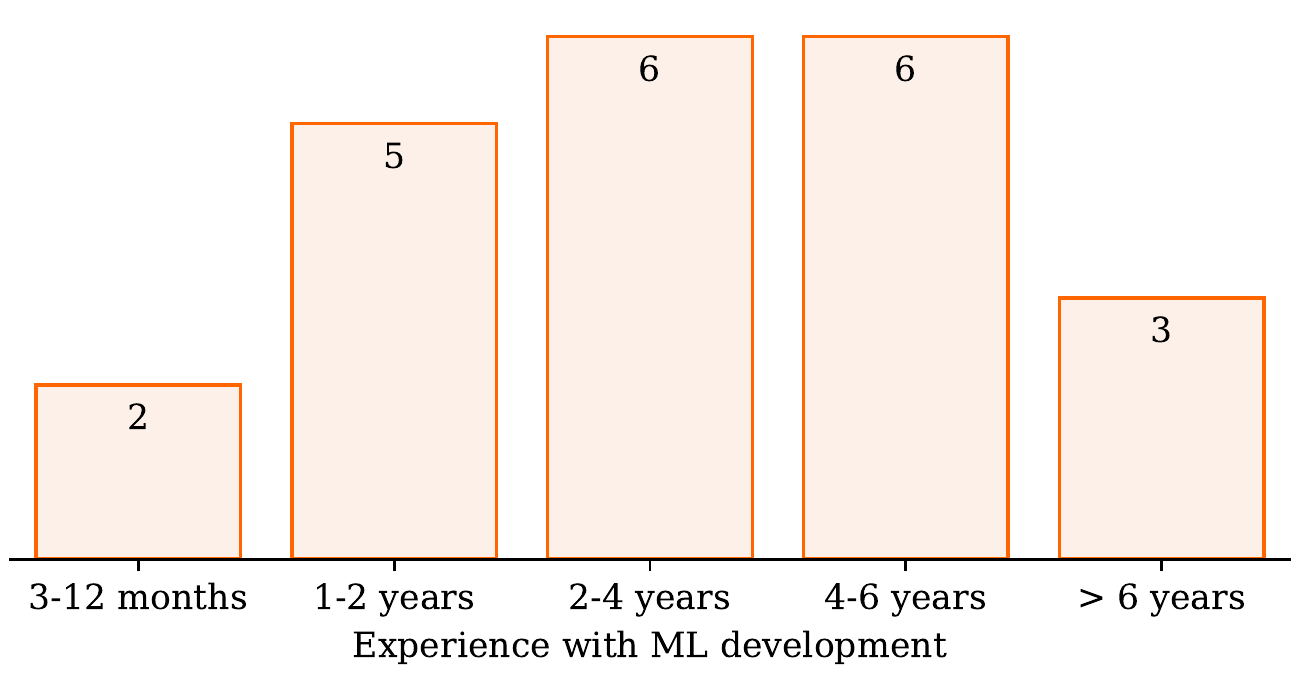}
    \vspace{-1em}
    \caption{Countplot of ML experience among our survey participants.}
    \label{fig:ml-experience}
    
    \includegraphics[width=0.8\linewidth]{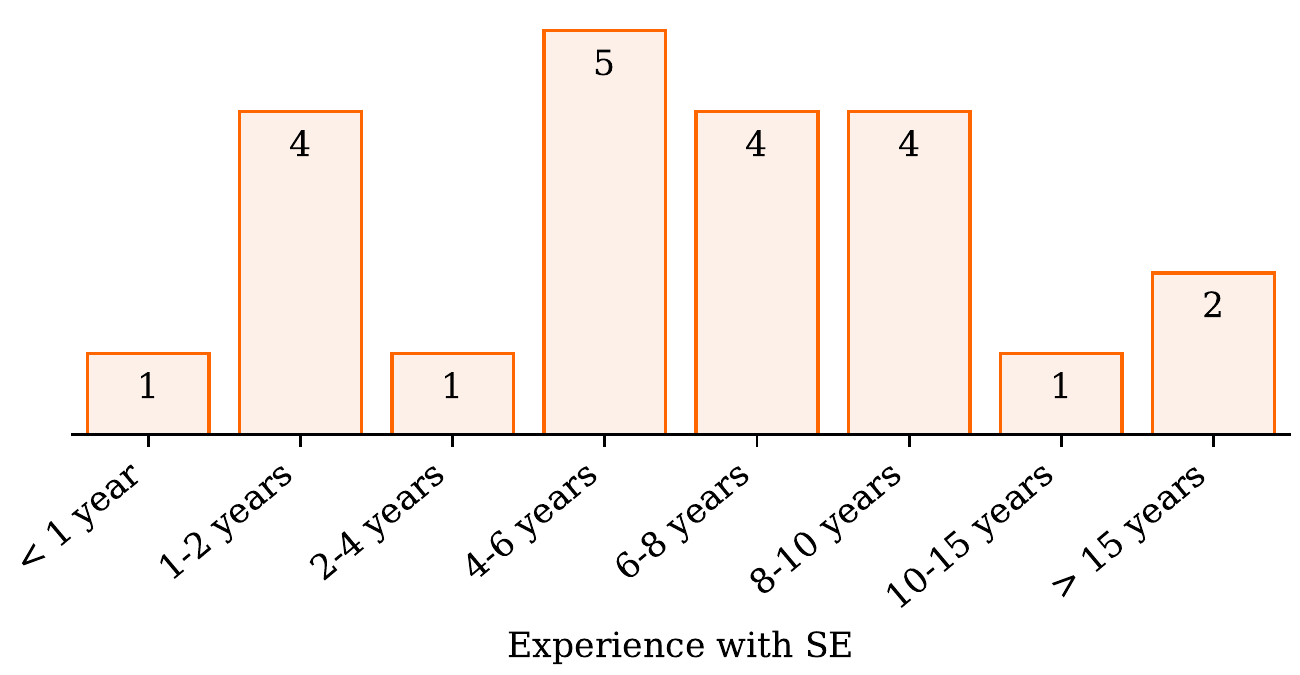}
    \vspace{-1em}
    \caption{Countplot of SE experience among our survey participants.}
    \label{fig:se-experience}
    
    \vspace{-1em}
\end{figure}



\section{Results} \label{sec:results}

Applying our methodology, we found the following answers to our research questions.

\subsection{RQ1: How do the project smells as detected by \mllint fit the industrial context of a large software- and data-intensive organisation like ING?}

In total, eight ML projects at ING were analysed. Four of these were proof-of-concept projects, two projects were production-ready, one project was in the process of being made production-ready and one project was an example project. Listing by \mllint category as described in \autoref{sec:implementation}, our key findings and observations are as follows\footnote{We omit the CI category, as its implementation in \texttt{mllint} has a false positive.}.

\textbf{Version Control}\\
Every project was using Git to version control their code. Three projects had large files in their Git history, some of it training data, some of it large Jupyter Notebook files. However, none of the projects that we analysed were version controlling their data using the Data Version Control (DVC)\footnote{\url{https://dvc.org/}} tool, though it is known that \textit{some} projects at ING do use it. Data acquisition methods differ per project: some receive it at run-time, one had scripts to retrieve the data from an external database, some instructed the user to download the data from an internal document sharing platform.
    
\textbf{Dependency Management}\\ 
Dependency management was done well in two projects, in one project not at all and in other projects with a combination of \codeinline{requirements.txt} and \codeinline{setup.py}, of which \mllint doesn't recognise whether it is used in an effective, maintainable and reproducible way. Manual inspection showed that these projects do groom their \codeinline{requirements.txt} files, there was no evidence of direct \codeinline{pip freeze} usage as was prevalent in \cite{bvo-wain2021} and some of these projects were neatly separating their runtime dependencies from development dependencies. However, there were also two projects that duplicated the contents of their \codeinline{requirements.txt} in their \codeinline{setup.py}.

\textbf{Code Quality}\\ 
The example project and (being made) production-ready projects adopt static analysis tools to lint for code smells, as indicated by instructions to run linters in the documentation or linter configurations in their repositories. These projects are not free of code smells though, as particularly Pylint was eager to complain, though it is disputable what degree of its messages were false positives or irrelevant. The other proof-of-concept projects, however, were not using static analysis tools, as shown by their lack of linter configuration, lack of linter usage instructions and abundance of detected code smells.

\textbf{Testing}\\ 
The two production-ready projects and example project have automated tests that also pass. Two proof-of-concept projects had varying amounts of tests, but some of them fail due to import errors\footnote{Note: this may also be caused by a misconfiguration on our end, though where available we did diligently follow instructions in the repository for running the tests.}. The other three projects, including the one being made production-ready, did not have any tests.

\subsection{RQ2: What are the differences between perceptions on \mllint's linting rules for proof-of-concept versus production-ready ML projects?}

From our survey, we have gathered the following results, as listed by linting rule (sub-)category. For the average importance, we encoded our five Likert-scale answers to integers between -2 and 2, and took the mean of the responses. For the range of importance, we subtracted and added the standard deviation from / to the mean, then rounded to the nearest integer, mapping back to a Likert-scale answer.

\textbf{Version Control -- Code}\\
For both proof-of-concept as well as production-ready projects, survey participants on average find usage of Git in ML projects between \textit{moderately important} and \textit{absolutely essential}, averaging \textit{very important}. For production-ready projects, usage of Git is even unanimously seen as \textit{absolutely essential}.

\textbf{Version Control -- Data}\\
The importance of the rules on the (correct) use of DVC is disputed: for proof-of-concept projects, survey participants find this between \textit{not important} and \textit{very important}, averaging to \textit{slightly important}. For production-ready projects, survey participants find this between \textit{slightly important} and \textit{absolutely essential}, averaging to \textit{very important}. Note, however, that the rules in this category primarily relate to the usage of the tool DVC, rather than the actual practice of version-controlling data, for which there exist many other options besides DVC.

\textbf{Dependency Management}\\
For proof-of-concept projects, the use of proper dependency management tooling is found to be between \textit{slightly important} and \textit{absolutely essential}, averaging \textit{very important}. While all other rules on proof-of-concept projects were lowest rated as \textit{not important}, this rule was the only rule to be lowest rated as \textit{slightly important}. For production-ready projects, this rule was rated between \textit{very important} and \textit{absolutely essential}, averaging \textit{absolutely essential}, with the lowest rating being \textit{moderately important}.

For both types of projects, however, the importances of using a single dependency manager and making a correct distinction between runtime and development dependencies, was disputed. The former was rated between \textit{slightly} vs. \textit{moderately important} and \textit{absolutely essential}, averaging \textit{moderately} vs \textit{very important}. The latter was rated between \textit{not} vs. \textit{slightly important} and \textit{very important} vs. \textit{absolutely essential}, averaging \textit{moderately important}.

\textbf{Continuous Integration}\\
For proof-of-concept projects, the use of CI was rated between \textit{slightly} and \textit{very important}, averaging \textit{moderately important}. For production-ready projects, this was rated between \textit{moderately important} and \textit{absolutely essential}, averaging \textit{very important}.

\textbf{Code Quality}\\
The recommendation to use code quality linters does see a significant shift in importance between proof-of-concept and production-ready projects. For a proof-of-concept project, our survey participants rate this between \textit{not important} and \textit{very important}, averaging \textit{moderately important}. For a production-ready project, they rate this between \textit{moderately important} and \textit{absolutely essential}, averaging \textit{very important}.

As for the actual linting tool being used, there is no significant difference in the perceived importance. There is a slight tendency towards the code formatting tool Black in proof-of-concept projects and towards the security-focused linter Bandit in production-ready projects. Overall, we find that the \textit{usage} of code quality linters is more important than a total absence of linter warnings.

\textbf{Testing}\\
The importance of having automated tests in a proof-of-concept ML project is disputed and perceived to be between \textit{slightly} and \textit{very important}, averaging \textit{moderately important}. For a proof-of-concept project, however, their importance is significantly higher, between \textit{moderately important} and \textit{absolutely essential}, averaging \textit{very important}. Passing the tests and having a test coverage report is also seen as \textit{moderately} vs. \textit{very important}.

\subsection{RQ3: What are the main obstacles for ML practitioners towards implementing specific best practices?}
Survey participants noted that out of \mllint's rules, they were most obstructed in implementing the practices about code quality linters and dependency management.

``\textit{Linters, specially regarding code quality, can be overwhelming if not properly configured. Not all warnings pointed are necessarily bad for your code, not all justified warnings are equally bad, so it needs to be used parsimoniously.}''

Regarding code quality, survey participants complained that linters generally suffer from a high degree of false positives and that configuring these linters is often cumbersome and time-consuming. They experience a catch-22 situation: on the one hand, using the default configuration leaving all rules enabled, in many cases results in an overwhelming amount of linter warnings that in the eyes of the user often do not relate to the project in a functionally meaningful way (e.g., trailing whitespace and proper docstring formatting, but also false positive type-checking errors). On the other hand, selectively enabling or disabling linting rules by configuring each linter for the project, is found to be time-consuming, difficult and cumbersome, especially for those inexperienced with the tool or the kind of linting rules and their importance to the quality of the project. This is especially found to be difficult in a team situation, as each developer may have different preferences / opinions about specific linting tools and rules.


Regarding dependency management, while \mllint recommends using Poetry or Pipenv, some survey participants note that they prefer to stick with Python's standard \codeinline{requirements.txt} and \codeinline{setup.py} files. While they do acknowledge that these are easy to misuse, especially for those inexperienced with them, several arguments are made for them. First, they argue that a disciplined developer or team can still use Python's standard tooling in an effective, maintainable and sufficiently reproducible manner, especially when combined with Docker. Secondly, if they are already sufficiently proficient at this, they do not want to re-learn to do with Poetry what they can already do with Pip. Thirdly, they note that they do not want to be forced to use external tooling (outside of Python) to interact with a project. Finally, they note that Poetry may conflict with other tools they are using in the project, such as Versioneer. Solving such conflicts creates extra overhead and may be further complicated by a smaller user base as opposed to standard Python tools to help with solving these conflicts.

Generalising with other mentions of less specific obstructions, we find that:

\begin{enumerate}
    \item Configuration of tools, especially static analysis tools, is perceived as difficult, cumbersome and time-consuming.
    \item False positives are a significant obstruction to the adoption of static analysis tools. They produce noise and may distract the user towards irrelevant or trivial issues, resulting in overhead to selectively ignore them or adjust the configuration of the tool to match their preferences.
    \item While some tools are recognised to be useful to inexperienced practitioners in reducing mistakes, experienced practitioners prefer to use tooling they are already used to, instead of having to learn a new tool. They may also be hesitant to add more tools to a tool stack they already deem sufficient.
    \item New tooling may conflict or have unexpected interactions with existing tooling. Solving these problems causes extra overhead for practitioners. The expectancy of such problematic interactions also creates apprehension towards adopting these tools.
\end{enumerate}



\subsection{RQ4: What is the perceived benefit of using static analysis tools such as \mllint to check SE guidelines in ML projects?}

Participants note that they find tools such as \mllint to be particularly useful for enforcing best practices, maintaining consistency in the project and discovering useful new tools in the Python ecosystem to improve their project quality. They also mention that static analysis tools such as \mllint can help guide the direction that further development efforts should take in the road towards productionisation of the project. In doing so, these tools' reports can be used as a project quality checklist.

``\textit{It is a great way to enforce best practices, avoid common pitfalls and automate "common sense". Without such tools it's easy to be sloppy on project quality.}''

Participants find static analysis tools such as \mllint to also be useful for doing code review by helping to ``\textit{bring attention of a reviewer to potentially problematic pieces of code introduced, specially when integrated to automated CI pipelines.}''. Usage of these tools in CI is a popular suggestion, though varying suggestions are made as to the frequency of running them (e.g. on every pull request, at every release, or before finalising the project).
\section{Discussion} \label{sec:discussion}

This section collects, combines and discusses our findings, sectioned along three themes: version controlling data, dependency management and static analysis tool adoption.

\subsection{Data version control}
While the results from RQ1 showed that none of the projects we analysed were using the tool DVC, this does not necessarily mean that industry ML practitioners are not version controlling their data. Validating with experienced ML engineers at ING, we find various ways in which data is managed: some projects only require data from the user at run-time, some have data small enough to fit in the code repository, but most prevalently, data is either pulled from a Hadoop filesystem or shared through other internal data sharing solutions, requiring the ML developer to download it manually. One ML engineer mentioned that they had experimented with DVC before, but found that it produced some overhead and preferred to stick with the semi-versioned workflow that they already had. Each of these methods has a varying degree of version control and a varying suitability towards certain types of data (consumption).

Overall, there seems to be a lack of standardised tooling for dealing with varying kinds of data dependencies, as practitioners generally stick with what is most practical or known to them. This presents a significant challenge to static analysis tools for detecting data version control techniques.

\subsection{Dependency management}
Overall, dependency management is perceived to be \textit{very important} vs. \textit{absolutely essential} and is done better in our industrial context than in the open-source context seen in~\cite{bvo-wain2021}, though practices still differ significantly between projects and developers. Some prefer to use external tooling such as Poetry, others prefer to create a manual workflow around Python's standard \codeinline{requirements.txt} and \codeinline{setup.py} files. Such a workflow was argued by several ML engineers to be effectively usable with disciplined and sufficiently experienced users, though they do recognise that they are prone to misuse with less experienced users.

This is particularly troubling in ML project development, given the gap in SE experience in data scientists. Unlike other popular language ecosystems such as NodeJS (npm or yarn), Go (built-in) and Rust (cargo), which external tools like Poetry take heavy inspiration from, the Python language ecosystem still lacks a standardised, easy-to-use, consistently used, maintainable and reproducible method of managing code dependencies.


\subsection{Static analysis tool adoption}
Results show a mixed sentiment towards static analysis tools. Combining findings from RQ1 and RQ2, we identify a tendency against using these linters during the development phase of the project and instead only adopting them during the productionisation phase, as an after-the-fact check on code quality. However, research has shown that linters are particularly useful during development for automatically fixing code styling (maintaining code consistency), avoiding complex code, and finding potential bugs early~\cite{tomasdottir2020eslint}. Especially in ML applications, where one run of the program could take hours, linters can save the user from an unfortunate typo in the program that would void all the time spent running it.


So why do practitioners refrain from adopting static analysis tools during their ML project development? Findings from RQ3, corroborated by existing research~\cite{tomasdottir2020eslint}, show two primary obstructions towards their adoption: a high rate of false positives and cumbersome, time-consuming configuration.

False positives in static analysis tools are particularly common in dynamically typed, interpreted languages such as Python. Our previous research also found Pylint to produce a high rate of false positives on imports, both local imports as well as prominent ML libraries~\cite{bvo-wain2021}. This problem is easier stated than solved, however, which calls for more development efforts and research into preventing false positives in static analysis on Python code, both from a tooling and linguistic perspective.

Practitioners could also selectively disable the rules that produce false positives in the configuration of their linters. However, as our findings corroborate, creating and maintaining linter configurations is also a significant challenge in their adoption~\cite{tomasdottir2020eslint}. Practitioners, especially those in a team and / or inexperienced with the linter or their importance to the code quality of the project, find it difficult, cumbersome and time-consuming to define their standards and configure their linters to fit them.

Thus, for tool developers to have their tools become widely used, there seems to be an inherent trade-off between having a tool that is abstract or malleable enough to fit as many applications as possible, while also requiring as little configuration effort from the user as possible. To achieve this, one set of recommendations is to simplify the configuration of the tool as much as possible: do not give the user unnecessary configuration options~\cite{xu2015too-much-configuration}. Similar to tools like \codeinline{gofmt}, \codeinline{govet} and Black: set standards and defaults that every user can agree on, then provide users with minimal knobs to adjust these, which barely leaves any room for bikeshedding.

However, many static analysis tools, including \mllint, are too complex or deal with too opinionated subjects to set \textit{one} standard. Thus, there is also a need for \textit{context-aware} static analysis tooling: by either automatically detecting or having the user configure in a simple way what the context of the project under review is, the tool can automatically adjust linting practices to conform to the user's needs. In the case of \mllint, this context would include the project's (desired) maturity (e.g. proof-of-concept, production-ready), use-case or tool stack. Context-aware static analysis functionality could be realised by using presets or profiles, similar to ESLint~\cite{tomasdottir2020eslint} and \codeinline{isort}, though more research on this subject is recommended.

\section{Threats to Validity} \label{sec:threats}

\subsection{Analysed projects \& survey participants}

In the qualitative analysis of \mllint reports of ING ML projects, we were only able to get access to \mllint reports of eight projects. There may also be a selection bias in the projects that we got access to, as we were unable to get access to ML projects of more sensitive natures, due to internal regulations around either what data they process or what functionality they perform. We are thus unsure of the generalisability of this dataset towards the state of ML projects at ING. We attempted to mitigate this problem by validating our generalised findings with experienced ML engineers at ING.


The limited number of survey participants (22) also presents a challenge to this research's validity. It was difficult to gain survey responses, in part also since it required participants to have experience running \mllint on one of their own ML projects. However, the more qualitative nature of our survey questions and analysis of the answers, combined with validation with experienced ML engineers, may help to mitigate this challenge.

We also find a lack in diversity among our survey participants, given that most of the participants are ML engineers with extensive experience in SE. While it is interesting to research how our tool relates to experienced engineers, it would be interesting to include more data scientists with little experience in SE in our survey.


\subsection{Construct validity}

For this research, we primarily focused on the project smells that \mllint is able to detect reliably. However, there are many more project smells to be catalogued and to be detected by \mllint. Additionally, there are many more tools, techniques and SE practices that \mllint does not yet recognise or recommend, but are valid ways of mitigating project smells. Some of \mllint's linting rules may currently also pertain more to the use of a tool that implements a certain ML project best practice, than to the practice itself, such as the rules about data version control with DVC. Therefore, until extended to support more, \mllint can only reliably detect a limited set of project smells, consequently limiting the scope of this research.

Furthermore, for the RQ4, we focus on \textit{perceived} benefits of static analysis tools, though it may be more interesting to investigate \textit{observable} benefits. This could be done, for example, by asking ML practitioners to use \mllint in the development of their ML projects for a prolonged period of time and then seeing how the software quality of their ML projects increases over time. While this was considered, we acknowledge that \mllint is not yet mature enough to accurately measure the full software quality of an ML project. To the best of our knowledge, there is also no other tool that can accurately measure the full software quality of an ML project, without limiting its scope to only one or a few aspects of software quality. This poses a challenge to future research on this subject.



\section{Conclusion \& Future Work} \label{sec:conclusion}
In conclusion, this research introduced the novel concept of \textit{project smells} as a more holistic view over code smells for assessing the software quality in ML projects and implemented a novel static analysis tool, \mllint, to help detect and mitigate these deficits in ML project management. We investigated: 1) how these project smells fit the industrial context of ING; 2) the perceived importance of \mllint's linting rules for ML practitioners on proof-of-concept versus production-ready projects; 3) the primary obstacles towards implementing solutions to these project smells and 4) the perceived benefits of using static analysis tools such as \mllint to verify SE practices in ML projects.


Future work should primarily focus on further development of \mllint, formally defining project smells and analysing 
their observable impact on the software quality of ML projects. Another interesting research direction is to investigate how ML development ecosystems can be redesigned or extended to inherently prevent such project smells from occurring.

Regarding the development of \mllint, more linting rules need to be implemented (particularly ML-specific rules), such as on asserting data quality, having a reproducible, end-to-end ML pipeline and linting for ML-specific code smells. Additionally, existing linting rules should be extended with more in-depth analysis of the project under review, with support for more toolsets and ML frameworks.

Future research could also investigate how \textit{context-aware} static analysis can be applied in practice and how existing static analysis tools can adopt this. For \mllint it would entail implementing methods for detecting or in a simple way configuring the technology stack or ML framework used in the project and the project's (desired) maturity. \mllint can then selectively disable non-applicable rules and adjusts the weights of other rules to fit with what is important for the project at that stage of development. However, the effort required to configure \mllint correctly, must also be minimal.

\bibliographystyle{ACM-Reference-Format}
\interlinepenalty=10000 
\bibliography{refs}

\appendix

\end{document}